\documentclass[preprint,showpacs,prb,amsmath,amssymb]{revtex4}

\usepackage{graphicx}
\usepackage{dcolumn}
\usepackage{bm}

\usepackage{natbib}

\begin{document}

\title{A detailed investigation of the properties of a Vlasov-Maxwell equilibrium for the force-free Harris sheet}
\author{T.Neukirch}
\email{thomas@mcs.st-and.ac.uk}
\author{F. Wilson}
\email{fionaw@mcs.st-and.ac.uk}
\author{M. G. Harrison}
\email{mikeh@mcs.st-and.ac.uk}

\affiliation{School of Mathematics and Statistics, University of St. Andrews, 
St. Andrews KY16 9SS, United Kingdom}

\begin{abstract}
A detailed discussion is presented of the Vlasov-Maxwell equilibrium for the force-free Harris sheet recently found by Harrison and Neukirch  (Phys. Rev. Lett. 102, 135003, 2009).  The derivation of the distribution function and a discussion of its general properties and their dependence on the distribution function parameters will be given. In particular,
the distribution function can be single-peaked or multi-peaked in two of the velocity components, with possible implications for stability. The dependence of the shape of the distribution function on the values of its parameters will be investigated and the relation to macroscopic quantities such as the current sheet thickness will be discussed.

\end{abstract}

\pacs{52.20.-j, 52.25.Xz, 52.55.-s, 52.65.Ff}

\maketitle

\section{Introduction}

Force-free plasma equilibria with the property
\begin{equation}
\mathbf{j} \times \mathbf{B}= \frac{1}{\mu_0}(\nabla \times \mathbf{B}) \times \mathbf{B} = \mathbf{0},
\label{forcefree1}
\end{equation}
are of great importance, in particular for space and astrophysical plasmas. 
Equation (\ref{forcefree1}) implies that the current density, $\mu_0 \mathbf{j} =\nabla \times \mathbf{B}$, is parallel to the magnetic field, $\mathbf{B}$, so that it can be written as $\mu_0 \mathbf{j} = \alpha \mathbf{B}$. In general, the function $\alpha$ can vary with position, but has to be constant along magnetic field lines, since $\nabla\cdot \mathbf{j} = 0$ together with $\nabla\cdot \mathbf{B} =0$ implies that 
\begin{equation}
\mathbf{B}\cdot \nabla \alpha =0.
\label{forcefree2}
\end{equation}
Obviously Eq. (\ref{forcefree2}) is also satisfied if $\alpha =$ constant. This case is usually referred to as the linear force-free case, because the equation determining the magnetic field is linear in this case. Magnetic fields for which $\alpha$ varies from field line to field line are called nonlinear force-free fields.

Whereas many force-free equilibria can be found using magnetohydrodynamics (MHD), this is not the case when Vlasov-Maxwell (VM) theory is used. Collisionless force-free equilibria have only been found for the special case where the magnetic field depends only on one spatial Cartesian coordinate (in this paper taken to be $z$). This case is trivial in MHD, but finding the appropriate distribution functions for given magnetic field and current density profiles for a collisionless equilibrium is a highly nontrivial  task. 
The reason for this difficulty is that one has to try and solve the VM problem in the opposite direction than it is usually treated, which is to specify the distribution functions (DFs) and then to calculate the magnetic field by solving Amp\`{e}re's law.

This difficulty is reflected by the fact that only a very small number of exact force-free VM equilibrium DFs are known and all known solutions were 
of the linear force-free type\cite{Sestero-1967,Channell-1976,Bobrova-1979,Bobrova-2001} until the first nonlinear force-free VM equilibrium DF was presented in a recent Letter.\cite{Harrison-2009b}
The DFs found in Ref.~\onlinecite{Harrison-2009b} are for the force-free Harris sheet, with a magnetic shear field ensuring force balance instead of a plasma pressure gradient as in the original Harris sheet.\cite{Harris-1962} 

For reasons of space no detailed discussion of  a) the derivation of the DFs and b) their properties has been given in Ref. \onlinecite{Harrison-2009b}. In the present publication, we aim to give a full discussion of the method used to derive the DFs in Sect. \ref{sec:calculation} and of its properties in Sect. \ref{sec:properties}. Of particular interest is the possibility that the DFs can have multiple maxima in two of the velocity components (in the coordinate system used in this paper the $v_x$- and $v_y$-components), which may have stability implications. Therefore, a detailed investigation of the connection between the shape of the distribution function and the parameter values was carried out. A summary and conclusion will be presented in Sect. \ref{sec:conclusions}.

\section{Calculation of the equilibrium distribution function}

\label{sec:calculation}

\subsection{Basics}

We use Cartesian coordinates $x$, $y$, $z$ complemented by the corresponding velocities 
$v_x$, $v_y$, $v_z$ for the DFs. We assume spatial invariance in $x$ and $y$, i.e. all quantities depend only upon $z$. We also assume time-independence. 

For the problems considered in the present paper the magnetic field has only two non-vanishing components, $B_{x}$ and $B_{y}$, which, using an appropriate gauge, can be written  in terms of a vector potential 
$\mathbf{A}=(A_x,A_y,0)$ in the form
\begin{equation}
  B_{x}=-\frac{dA_{y}}{dz},   
  \label{bx}
\end{equation}
\begin{equation}
  B_{y}=\frac{dA_{x}}{dz}.
  \label{bz}
\end{equation}
The electric field is given by the negative gradient of an electric potential $\phi$ such that
\begin{equation}
  \mathbf{E}=-\nabla\phi=-\frac{d\phi}{dz}\mathbf{e_{z}} .
  \label{efield}
\end{equation}
The magnetic and electric fields thus automatically satisfy the homogeneous steady-state Maxwell equations $\nabla \cdot \mathbf{B}=0$ and $\nabla \times \mathbf{E} =\mathbf{0}$.

Due to time independence and spatial symmetries we have three obvious constants of motion for particles of species $s$ with charge $q_s$ and mass $m_s$ moving in these fields, namely
the particle energy, $H_s$,
\begin{equation}
H_{s}=\frac{1}{2} m_{s} (v_x^2+v_y^2+v_z^2)+q_{s}\phi,
  \label{hamiltonian}
\end{equation}
the canonical momentum in the $x$-direction, $p_{xs}$,
\begin{equation}
  p_{xs}=m_{s}v_{x}+q_{s}A_{x},
  \label{px}
\end{equation}
and the canonical momentum in the $y$-direction, $p_{ys}$,
\begin{equation}
  p_{ys}=m_{s}v_{y}+q_{s}A_{y}.
  \label{py}
\end{equation}

Solutions of the steady state Vlasov equation 
\begin{equation}
  \mathbf{v} \cdot \frac{\partial f_{s}}{\partial \mathbf{r}}+ \frac{q_s}{m_s}(\mathbf{E+v\times B}) \cdot \frac{\partial f_{s}}{\partial \mathbf{v}}=0.
  \label{steadystatevlasov}
\end{equation}
are given by
all positive functions $f_s $ 
depending only on the constants of motion,
\begin{equation}
  f_{s}=f_{s}(H_{s},p_{xs},p_{ys}),
  \label{equilibriumf}
\end{equation}
and satisfying the appropriate conditions for existence of the velocity moments.
If the same combination of values for the constants of motion allows particle trajectories in several distinct regions of phase space then it is in principle possible to assign different values to 
$f_s$ in each region\cite{Grad-1961,Mynick-1979a}, but this possibility will not  be considered in the present paper  (for an example of 2D rotationally symmetric VM equilibria see e.g. Ref. \onlinecite{Neukirch-1993}).

Using the assumption of quasineutrality to determine the electric potential $\phi$, one can show\cite{Mynick-1979a,Harrison-2009a} that the VM equilibrium problem reduces to solving Amp\`{e}re's law in the form
\begin{eqnarray}
  \frac{d^{2}A_{x}}{dz^2} &= &-\mu_0 j_x= -\mu_{0}\frac{\partial P_{zz}}{\partial A_{x}} ,\label{amp1}\\
  \frac{d^{2}A_{y}}{dz^2}&=&-\mu_0 j_y= -\mu_{0}\frac{\partial P_{zz}}{\partial A_{y}} ,\label{amp2} 
  \end{eqnarray}
where
\begin{equation}
P_{zz}(A_x,A_y) = \sum_s m_s \int \,v_z^2 f_s \, d^3v
\label{pzzdef}
\end{equation}
is the $zz$-component of the plasma pressure tensor.

It is obvious (see e.g. Ref \onlinecite{Harrison-2009a}) that Eqs. (\ref{amp1}) and (\ref{amp2}) are equivalent to the equations of motion of a particle in a 2D conservative potential, with $z$ taking the role of time, $A_x$ and $A_y$ the coordinates of the particle and $\mu_0 P_{zz}$ being the potential. As in the analogous particle problem one can integrate Eqs. (\ref{amp1}) and (\ref{amp2}) once to get
\begin{equation}
\frac{d}{dz} \left[ \frac{1}{2\mu_0} \left( \frac{ d A_x}{d z} \right)^2  + 
\frac{1}{2\mu_0} \left( \frac{d A_y}{dz} \right)^2 + P_{zz}(A_x,A_y) \right] = 0
\end{equation}

so that
\begin{equation}
 \frac{1}{2\mu_0} \left( \frac{ d A_x}{d z} \right)^2  + 
\frac{1}{2\mu_0} \left( \frac{d A_y}{dz} \right)^2 + P_{zz}(A_x,A_y) =  P_{total} =\mbox{ constant},
\label{totalpressureconst}
\end{equation}
i.e. the total pressure (magnetic plus plasma pressure) is constant for this class of VM equilibria.
The total pressure corresponds to the total energy in the particle problem. 

Knowledge of the shape of $P_{zz}(A_x,A_y)$ can provide insight into the nature 
of the solutions of Eqs. (\ref{amp1}) and (\ref{amp2}) in the same way as knowledge of the potential as a function of position in the equivalent particle problem can provide information about the nature of the possible trajectories of the particle. It is usually straightforward to calculate $P_{zz}$ as a function of $A_x$ and $A_y$ if the equilibrium DFs are specified. It may, however, also be possible to determine equilibrium DFs for a given function $P_{zz}(A_x,A_y)$ using a 
method suggested by Channell.\cite{Channell-1976}

\subsection{Channell's Method}

To be able to make analytical progress in determining a distribution function from $P_{zz}(A_x,A_y)$ a number of assumptions have to be made. The first assumption made is that the dependence of the DFs on the Hamiltonian $H_s$ is of the form of a negative exponential, i.e
\begin{equation}
f_s(H_s, p_{xs}, p_{ys}) = \frac{n_{0s}}{(\sqrt{2\pi}v_{th,s})^3}\exp(-\beta_s H_s) g_s(p_{xs}, p_{ys}) 
\label{channellass1}
\end{equation}
with $\beta_s = (k_B T_s)^{-1}$, $v_{th,s} = (\beta_s m_s)^{-1/2}$and $g_s$ an unknown function of the canonical momenta. Using this DF $P_{zz}$ becomes
\begin{equation}
P_{zz} = \sum_s \frac{1}{ \beta_s }  \exp(-\beta_s q_s \phi) N_s(A_x,A_y),
\label{pzzchannellgeneral}
\end{equation}
with
\begin{equation}
N_s(A_x,A_y)  = \frac{n_{0s}}{2 \pi v_{th,s}^2 }
\int\limits_{-\infty}^\infty \int\limits_{-\infty}^\infty  
\exp\left[-\frac{\beta_sm_s}{2}( v_x^2 +v_y^2 )\right]
g_s(p_{xs}, p_{ys} ) \, dv_x dv_y.
\label{channeldensitygeneral}
\end{equation}
The charge density, $\sigma$, can be calculated by taking the negative derivative of $P_{zz}$ with respect to the electric potential\cite{Mynick-1979a,Harrison-2009a} as
\begin{equation}
\sigma(A_x,A_y,\phi) = \sum_s q_s \exp(-\beta_s q_s \phi) N_s(A_x,A_y).
\label{chargedensity}
\end{equation}
Assuming a two-species plasma with both species having the same charge $e$ with 
opposite sign (e.g. electrons and protons) and quasi-neutrality 
by letting $\sigma =0$, one can determine the quasi-neutral electric potential to be
\begin{equation}
\phi_{qn} = \frac{1}{e(\beta_e+\beta_i)} \ln(N_i/N_e).
\end{equation}
Channell's\cite{Channell-1976} final assumption is strict neutrality, i.e. that $N_i (A_x,A_y)= N_e(A_x,A_y)=N(A_x,A_y)$ 
for all possible values of $A_x$, $A_y$, implying that $\phi_{qn}=0$. This will impose additional conditions on the parameters of the DFs which have to be satisfied, but this is in principle not a problem.

The neutral $P_{zz}$ is then given by
\begin{equation}
P_{zz}(A_x, A_y)  = \frac{\beta_e+\beta_i}{\beta_e \beta_i}N(A_x,A_y).
\label{pzzneutral}
\end{equation}
Using the canonical momenta instead of the velocity components as integration variables and 
using Eq. (\ref{pzzneutral}), 
Eq. (\ref{channeldensitygeneral}) becomes
\begin{eqnarray}
\frac{n_{0s}}{2 \pi m_s^2 v_{th,s}^2 }\int\limits_{-\infty}^\infty \int\limits_{-\infty}^\infty  
\exp\left\{-\frac{\beta_s}{2m_s}[ (p_{xs}- q_s A_x)^2 +(p_{ys}- q_s A_y)^2 ]\right\}
g_s(p_{xs}, p_{ys} ) \, dp_{xs} dp_{ys} = && \nonumber \\
&& \mbox{\hspace{-0.5\textwidth}} \frac{\beta_e \beta_i}{\beta_e+\beta_i} P_{zz}(A_x, A_y).
\label{channellfundamentaleq}
\end{eqnarray}
For $P_{zz}(A_x,A_y)$ a known function of $A_x$ and $A_y$, this is a Fredholm integral equation of the first type for the unknown function $g_s(p_{xs} , p_{ys})$. The kernel 
$K(p_{xs} , p_{ys}; q_s A_x, q_s A_y)$ of this integral equation 
\begin{equation}
K(p_{xs} , p_{ys}; q_s A_x, q_s A_y) \propto 
\exp\left\{-\frac{\beta_s}{2m_s}[ (p_{xs}- q_s A_x)^2 +(p_{ys}- q_s A_y)^2 ]\right\}
\end{equation}
depends only upon the difference of its arguments and the standard method for solving such integral equations is using Fourier transforms, as also suggested by Channell.\cite{Channell-1976}

It must, however, be pointed out that to be able to determine $g_s$ by Fourier transforms two conditions need to be satisfied: a) the Fourier transform of $P_{zz}(A_x, A_y)$ must exist and b) the inverse Fourier transform to obtain $g_s$ must exist. Especially condition b) can prove difficult to meet as the inverse Fourier transform involves a factor with the inverse of the Gaussian in the convolution integral, i.e. an exponential function with a \emph{positive} quadratic argument. Channell\cite{Channell-1976} treats several examples for which the Fourier transform method does not work using other methods. For the force-free Harris sheet case discussed in the present paper, we will also use a more direct method to 
solve Eq. (\ref{channellfundamentaleq}) because Fourier transforms are only of limited applicability to our case and because the other method turns out to be more instructive.

\subsection{Harris sheet and force-free Harris sheet}

The Harris sheet\cite{Harris-1962} is a well-known one-dimensional VM equilibrium. It is widely used in theoretical plasma physics, for example for reconnection studies, because it is a typical neutral current sheet and is mathematically well-behaved. The magnetic field is given by
\begin{equation}
\mathbf{B}_{Harris} = B_0 (\tanh(z/L), 0, 0),
\label{harrisB}
\end{equation}
the current density by
\begin{equation}
\mu_0 \mathbf{j}_{Harris} =B_{0}/L (0, 1/ \cosh^2(z/L),  0),
\label{harrisj}
\end{equation}
and the vector potential (in a convenient gauge) by
\begin{equation}
\mathbf{A}_{Harris} =B_{0} L (0, - \ln[\cosh(z/L)], 0).
\label{harrisA}
\end{equation}
Force balance is maintained by a pressure gradient with $P_{zz}(z)$ given by
\begin{equation}
P_{zz, Harris} = \frac{P_{0,zz}}{\cosh^2( z/l)}  + P_{b,zz},
\label{harrisP}
\end{equation}
with $P_{0,zz} = B_{0}^2/(2\mu_0)$ and $P_{b,zz}$ a constant background pressure.
The distribution function used by Harris\cite{Harris-1962} is 
given by
\begin{equation}
f_{s,Harris} =  \frac{n_{0s}}{(\sqrt{2\pi}v_{th,s})^3}\exp[-\beta_s ( H_s - u_{ys} p_{ys})],
\label{harrisdf}
\end{equation}
which is a Maxwellian DF in all velocity directions, but with a constant average bulk flow velocity of $u_{ys}$ in the $y$-direction. Other distribution functions giving rise to the same magnetic field and pressure profiles have also been found (see e.g. Ref.~\onlinecite{Fu-2005}).
By using either the distribution function (\ref{harrisdf}) directly or Eqs. (\ref{harrisA}) and (\ref{harrisP}), one can show that
\begin{equation}
P_{zz,Harris}(A_x,A_y) = P_{0,zz} \exp[ 2 A_y/(B_{0} L)] + P_{b,zz}.
\label{harrisPofA}
\end{equation}
Note that to get a constant background pressure from the distribution function an extra term 
proportional to
$\exp(-\beta_s H_s)$ has to be added to the right-hand side of Eq. (\ref{harrisdf}). Using that
$\tanh^2 x = 1 - 1/\cosh^2x$ the equilibrium condition (\ref{totalpressureconst}) is satisfied with
\begin{equation}
P_{total, Harris} = \frac{B_{0}^2}{2\mu_0 } + P_{b,zz}.
\label{harristotalpressure}
\end{equation}

\begin{figure}
\includegraphics[width=0.4\textwidth]{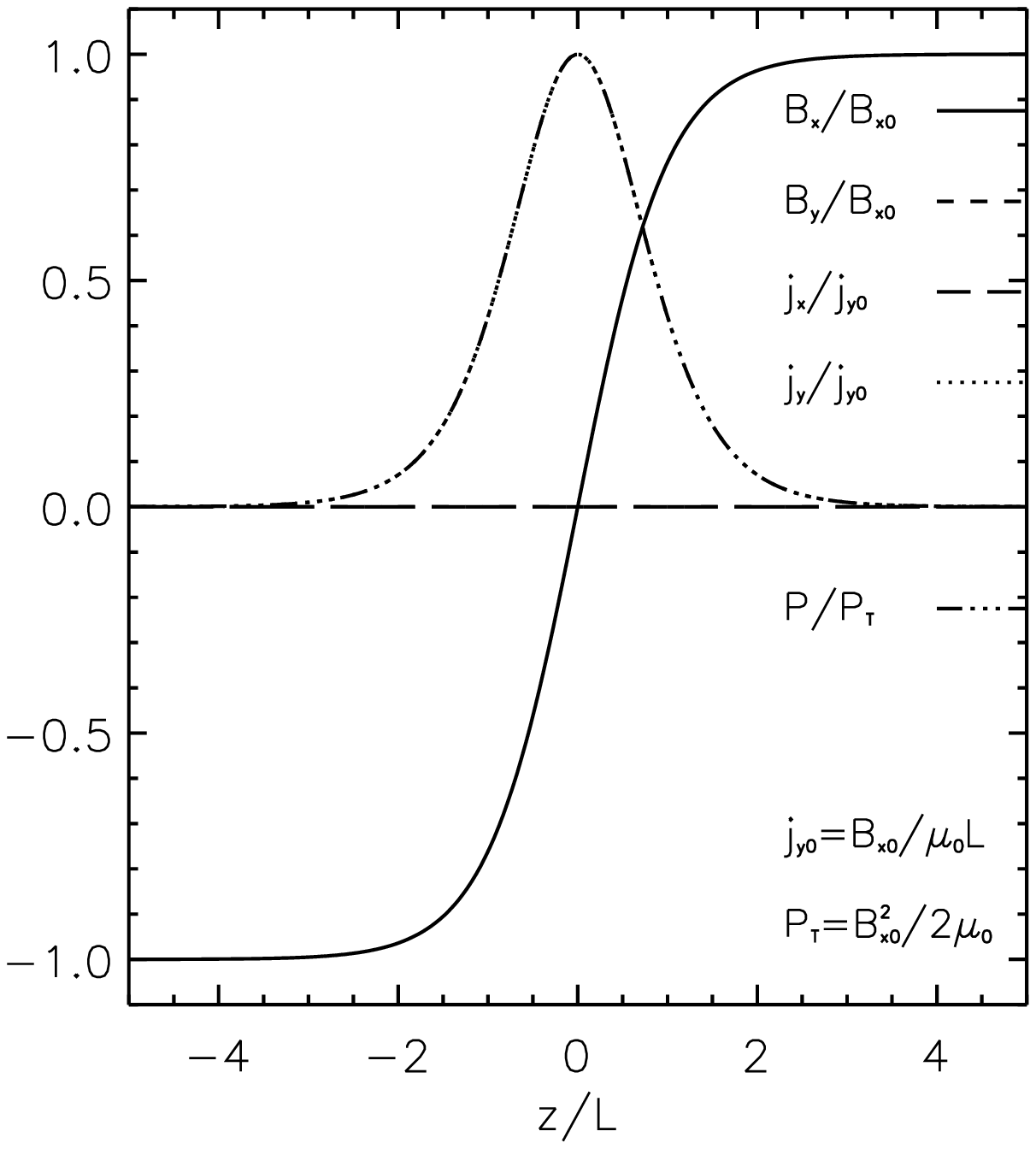}
\includegraphics[width=0.4\textwidth]{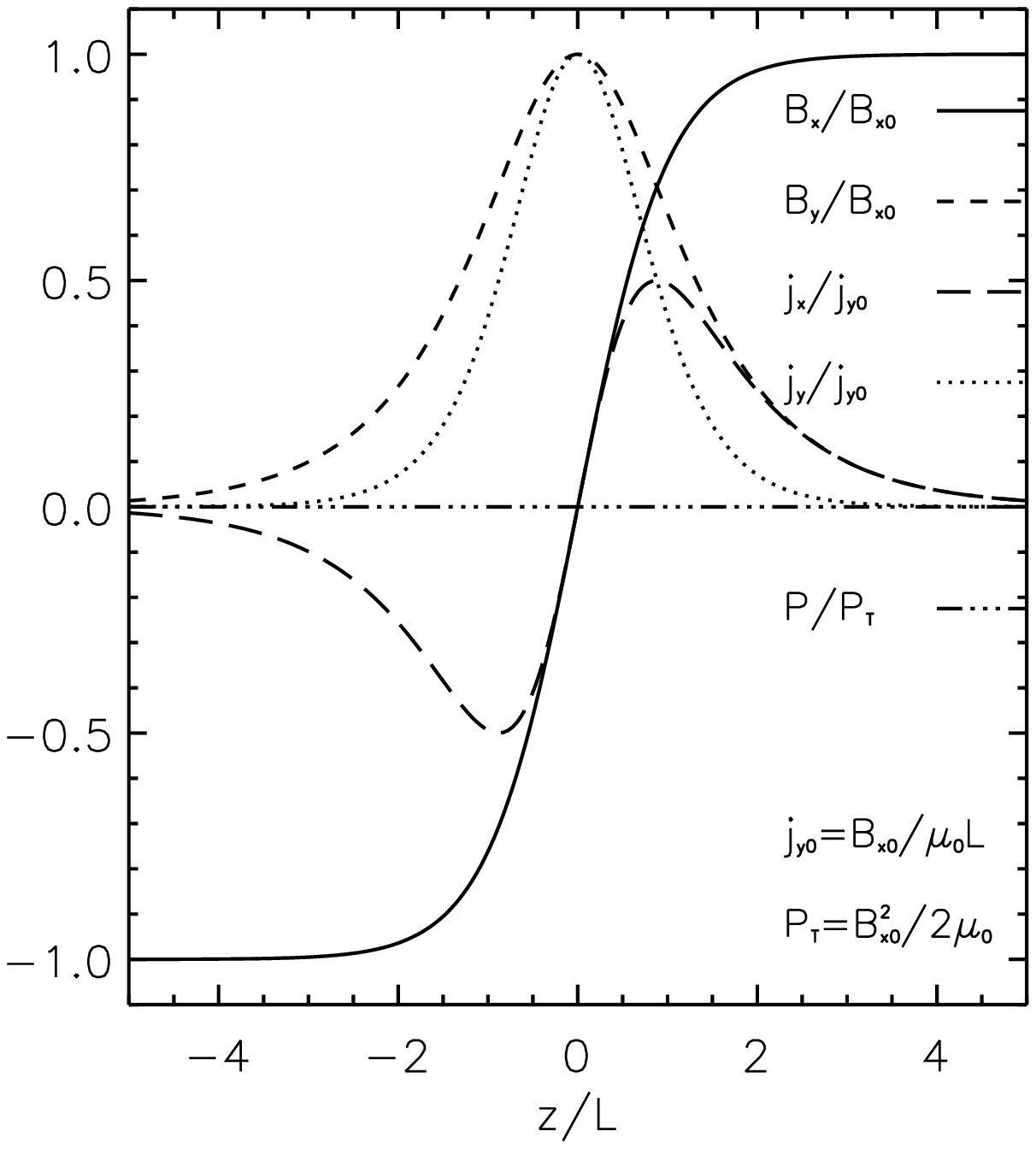}
\caption{The magnetic field, current density and pressure profiles as functions 
of $z/L$ for the Harris sheet (left panel) and the force-free Harris sheet (right panel).}
\label{fig:plotsofz}.
\end{figure}

The force-free Harris sheet has the same $B_x$ as the Harris sheet, but is kept in force balance by 
magnetic pressure due to a $B_y$ component, with $P_{zz}$ being constant. The magnetic field is then given by
\begin{equation}
\mathbf{B}_{ffHarris} = B_{0}(\tanh(z/L), 1/\cosh(z/L),0).
\label{ffharrisB}
\end{equation}
The current density is
\begin{equation}
\mu_0\mathbf{j}_{ffHarris} = B_{0}/L(\tanh(z/L)/\cosh(z/L),1/ \cosh^2(z/L),  0),
\label{ffharrisj}
\end{equation}
with $\mu_0\mathbf{j}_{ffHarris} = \alpha \mathbf{B}_{ffHarris}$ where
\begin{equation}
\alpha(z) = \frac{1}{L\cosh(z/L)}.
\label{alphaffharris}
\end{equation}
The vector potential, again in a convenient gauge, is given by
\begin{equation}
\mathbf{A}_{ffHarris} = B_0 L(2 \arctan(\exp(z/L)), -\ln(\cosh(z/L)),0).
\label{ffharrisA}
\end{equation}
At this point, no form for $P_{zz}$ as a function of $A_x$ and $A_y$ and no DF are known for this equilibrium magnetic field. The derivation of both will be discussed in the next section.
Plots of the magnetic field components, current density and pressure as functions of $z/L$ are shown in
Fig. \ref{fig:plotsofz}.

\subsection{Derivation of the distribution function}

To be able to apply Channell's method to find a DF for the force free Harris sheet, we first need to find an appropriate function $P_{zz}(A_x, A_y)$ for these cases.
It can be shown\cite{Harrison-2009a} that to find a $P_{zz}$ that allows a force-free solution is equivalent to finding a potential for which at least one of its equipotential lines is identical with a particle trajectory. The simplest examples for this are attractive central potentials whose contours are circles and which also allow circular orbits. The corresponding 1D VM equilibria are linear force-free magnetic fields.\cite{Sestero-1967,Channell-1976,Bobrova-1979,Bobrova-2001}

\begin{figure}
\includegraphics[width=0.8\textwidth]{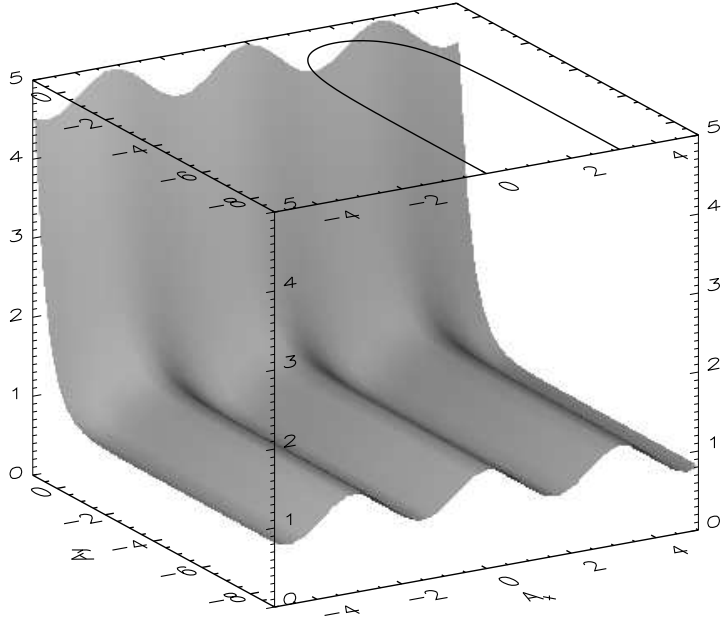}
\caption{Surface plot over the $A_x$-$A_y$-plane of the pressure function $P_{zz}(A_x,A_y)$ for the force-free Harris sheet. The vector potential of the force-free Harris sheet traces out a trajectory in the $A_x$-$A_y$-plane, which is shown at the top of the plot. This trajectory coincides with a 
contour of $P_{zz}(A_x,A_y)$, as the general condition for force-free VM 
equilibria demands.\cite{Harrison-2009a}}
\label{fig:pzzffharris}
\end{figure}

It is obvious, however, that the $P_{zz}$ for the force-free Harris sheet has to be more complex than a central potential. The approach chosen in Ref. \onlinecite{Harrison-2009b} was to let
\begin{equation}
P_{zz}(A_x,A_y) = P_1(A_x) + P_2(A_y).
\label{Pasasum}
\end{equation}
In this case the Eqs. (\ref{amp1}) and (\ref{amp2}) decouple and can be integrated separately. Thus one can see immediately that $P_2(A_y)$ is identical to $P_{zz,Harris}$ given by Eq. (\ref{harrisPofA}). The unknown function $P_1(A_x)$ can be determined from inverting $A_{x,ffHarris}(z)$ using 
Eq. (\ref{ffharrisA}) and substituting $z(A_{x,ffHarris})$ into 
\begin{equation}
P_1(z) = P_{1,b} - \frac{ B_0^2} {2 \mu_0} \frac{1}{\cosh^2(z/L)}.
\end{equation}
Using the trigonometric identity 
\begin{equation}
\sin(2x) = \frac{2 \tan x}{1+\tan^2 x},
\end{equation}
one can see that
\begin{equation}
\sin\left( \frac{ A_{x,ffHarris}}{B_0 L } \right) =\frac{1}{\cosh(z/L)},
\end{equation}
so that, dropping the subscript $ffHarris$,
\begin{equation}
P_1(A_x) = P_{b,1} - \frac{ B_0^2} {2 \mu_0}\sin^2\left( \frac{ A_{x}}{B_0 L } \right).
\end{equation}
Using $\sin^2 x = [1- \cos(2x)]/2$ and putting together $P_1(A_x)$ and $P_2(A_y)$,
we arrive at the form of $P_{zz}(A_x,A_y)$ given in Ref. \onlinecite{Harrison-2009b}
\begin{equation}
P_{zz}(A_x,A_y) = \frac{B_0^2}{2\mu_0} \left[ \frac{1}{2} \cos\left(  \frac{ 2 A_{x}}{B_0 L }  \right) +
\exp\left( \frac{2 A_y}{B_0 L}\right) \right] + P_{b},
\label{ffharrisPofA}
\end{equation}
where $P_{b} = P_{b,1} +P_{b,zz} - B_0^2/(4 \mu_0)$.
By construction, Amp\`{e}re's law (\ref{amp1}) and (\ref{amp2}) generated from this $P_{zz}$ has the vector potential (\ref{ffharrisA}) as a solution, and this solution coincides with a contour of 
$P_{zz}(A_x,A_y)$. In Fig. \ref{fig:pzzffharris} we show a surface plot of
$P_{zz}(A_x,A_y)$ for the force-free case with the vector potential for the force-free Harris sheet shown as a trajectory at the top of the plot.

Having found a $P_{zz}(A_x,A_y)$, we can use Channell's method\cite{Channell-1976} to find the corresponding DF. As the relation between the unknown function $g_s(p_{xs}, p_{ys})$ and $P_{zz}$
is linear, it is immediately clear that  $g_s$ must also have the form of a sum,
\begin{equation}
g_s(p_{xs}, p_{ys}) = g_{s1}(p_{xs}) + g_{s2}(p_{ys}),
\end{equation}
with
\begin{eqnarray}
\frac{\beta_e\beta_i}{\beta_e+\beta_i}P_1(A_x) & =& \sqrt{\frac{\beta_s}{2\pi m_s}} n_{0s}
 \int\limits_{-\infty}^\infty \exp\left[-\frac{\beta_s}{2 m_s} (p_{xs}-q_s A_x)^2\right] 
 g_{s1}(p_{xs}) \, dp_{xs} ,  
 \label{channell1}\\
\frac{\beta_e\beta_i}{\beta_e+\beta_i}P_2(A_y) & =&  \sqrt{\frac{\beta_s}{2\pi m_s}} n_{0s}
 \int\limits_{-\infty}^\infty \exp\left[-\frac{\beta_s}{2 m_s} (p_{ys}-q_s A_y)^2\right] 
 g_{s2}(p_{ys}) \, dp_{ys} . 
 \label{channell2}
\end{eqnarray}
For the time being we can ignore any constant parts of $P_1$ and $P_2$, because the solution for a constant $P$ is simply a constant $g$, which can be added at the end of the calculation due to the linearity of the problem.

For solving  Eq. (\ref{channell1}) with $P_1(A_x) \propto \cos(2A_x/B_0L)$ one could in principle use Fourier transforms, but we shall use a more direct method here. The method is based on the observation that, using the trigonometric identity $\cos[b(s+t)] = \cos(bs)\cos(bt) - \sin(bs) \sin(bt)$, one has
\begin{equation}
\int\limits_{-\infty}^\infty \exp(-a s^2) \cos[b(s+t)] ds =\sqrt{\frac{\pi}{a}} \exp\left( -\frac{b^2}{4a}\right)\cos(bt).
\end{equation}
Thus, rewriting the integral in Eq. (\ref{channell1}) using $v_x$ as integration variable 
instead of $p_{xs}$, one can see immediately that a $g_s(p_{xs}) \propto \cos(\beta_s u_{xs} p_{xs})$
leads to a $P_1(A_x) \propto \cos(\beta_s u_{xs} q_s A_x)$. The constant $u_{xs}$ has the 
dimensions of a velocity so that the argument of the cosine function is dimensionless.

The solution to Eq. (\ref{channell2}) is already known, because this part of the pressure gives rise to the $y$-component of the current density and thus to the Harris sheet $B_x$. Therefore, we must have  $g_{s2}(p_{ys} )\propto \exp(\beta_s u_{ys} p_{ys})$ (note that the case of a simple 
exponential $P_2(A_y)$ is also a special case of one the examples in Channell's paper\cite{Channell-1976}). 
This means that the part of the DF depending explicitly on $p_{ys}$ is identical with the $p_{ys}$-dependent part of the original 
Harris sheet DF (\ref{harrisdf}).

The full distribution function therefore has the general form
\begin{equation}
f_s = \frac{n_{0s}}{(\sqrt{2\pi}v_{th,s})^3} \exp(-\beta_s H_s) \left[a_s \cos \left(\beta_s u_{xs}p_{xs} \right) +  \exp \left( \beta_s u_{ys} p_{ys} \right)
   + b_s \right],
\label{fullDF}
\end{equation}
with $a_s$, $b_s$, $u_{xs}$ and $u_{ys}$ being constant parameters of the DF in addition to $n_{0s}$
and $\beta_s$. We remark  that we assume that $b_s > |a_s| \ge 0$ at this point to ensure that $f_s$ remains positive.
The parameters of the DF will have to satisfy a number of consistency relations due to the assumptions made for applying Channell's method and in order to relate the microscopic DF parameters to the macroscopic parameters $B_0$ and $L$.

\subsection{Consistency Relations}

The pressure tensor component $P_{zz}$ we obtain using Eq. (\ref{fullDF}) is of the general form
(\ref{pzzchannellgeneral}) with
\begin{eqnarray}
N_s(A_x,A_y) &= & n_{0s}  \exp\left(\frac{\beta_s m_s u_{ys}^2}{2}\right)  \left[
a_s\exp\left(-\frac{\beta_s m_s (u_{xs}^2+u_{ys}^2)}{2}\right)\cos(\beta_s u_{xs} q_s A_x)  \right.
 \nonumber\\
& &\qquad \qquad \left.+  \exp(\beta_s u_{ys} q_s A_y) + 
b_s \exp\left(-\frac{\beta_s m_s u_{ys}^2}{2}\right) \right].
\label{Nffharris}
\end{eqnarray}
The fundamental condition for Channell's method to be applicable is $N_e(A_x,A_y) = N_i(A_x,A_y)$.
This is satisfied if 
\begin{eqnarray}
n_{0e}  \exp\left(\frac{\beta_e m_e u_{ye}^2}{2}\right) & =& 
n_{0i}  \exp\left(\frac{\beta_i m_i u_{yi}^2}{2}\right)  = n_0 \label{defn0} \\
a_e \exp\left(-\frac{\beta_e m_e (u_{xe}^2+u_{ye}^2)}{2}\right) &=& 
a_i \exp\left(-\frac{\beta_i m_i (u_{xi}^2+u_{yi}^2)}{2}\right) = a \label{defa} \\
b_e \exp\left(-\frac{\beta_e m_e u_{ye}^2}{2}\right) &= &
b_i \exp\left(-\frac{\beta_i m_i u_{yi}^2}{2}\right) = b \label{defb} \\
\beta_e |u_{xe} | &=& \beta_i |u_{xi}|    \label{uxcondition} \\
-\beta_e u_{ye} &=& \beta_i u_{yi}  \label{uycondition} 
\end{eqnarray}
For the case of the original Harris sheet,  Eq. (\ref{uycondition}) is well known\cite{Schindlerbook} as the condition for a vanishing electric potential. In the Harris sheet case, $u_{ys}$ is the constant average 
bulk velocity of species $s$ in the $y$-direction and condition (\ref{uycondition}) is basically specifying a particular frame of reference. In the case of the force-free Harris sheet, the
average bulk velocity for both the $x$- and the $y$-velocity components varies with $z$ and one thus needs more conditions, but in principle
one can still interpret Eqs. (\ref{defn0}) to (\ref{uycondition}) as conditions for a particular frame of reference in which the electric potential vanishes.

Using Eqs. (\ref{defn0}) to (\ref{uycondition}) the general expression for $P_{zz}(A_x, A_y)$ for the force-free Harris sheet equilibrium becomes
\begin{equation}
P_{zz}(A_x, A_y) = \frac{\beta_e+\beta_i}{\beta_e\beta_i} n_0\left[
a\cos(e\beta_eu_{xe}A_x) + \exp(-e \beta_e u_{ye} A_y) +b
\right],
\label{pzzmicro}
\end{equation}
where, for simplicity, we have used the electron parameters only at the moment. An expression which is
symmetrical in the electron and ion parameters will be derived in Sect. \ref{sec:micromacro}.

\section{Properties of the equilibrium distribution function}

\label{sec:properties}

\subsection{Relation between microscopic and macroscopic parameters}

\label{sec:micromacro}

Although we have now derived the DF for the force-free Harris sheet, 
we have not yet related the set of microscopic parameters of the DF, namely $\beta_s$, $u_{xs}$,
$u_{ys}$, $a_s$ and $b_s$, to the macroscopic parameters of the equilibria, which are $B_0$ and $L$. The easiest way to find this connection is to compare Eq. (\ref{pzzmicro}) with Eq. (\ref{ffharrisPofA}).
This leads to 
\begin{eqnarray}
\frac{B_0^2}{2 \mu_0} &=& \frac{\beta_e + \beta_i}{\beta_e \beta_i} n_0, \label{b0n0} \\
\frac{1}{2}   & =& a, \label{ra}\\
P_b & =& \frac{\beta_e + \beta_i}{\beta_e \beta_i} n_0 b, \label{pbb} \\
\frac{2}{|B_0| L } &=& e \beta_e |u_{xe}| = e \beta_i |u_{xi}| ,\label{b0lux}\\
\frac{2}{B_0 L}  & = & -e \beta_e u_{ye} = e \beta_i u_{yi}, \label{b0luy}
\end{eqnarray}
where we have assumed  that $L$ is positive, but allow for $B_0$ to be negative.

To make the connection with the original Harris sheet results 
we use Eqs. (\ref{b0luy}) and (\ref{b0n0}) to derive an expression for $L$ in the form 
(see also Ref. \onlinecite{Schindlerbook}, Chapter 6)
\begin{equation}
L =\left( \frac{2(\beta_e+\beta_i)}{\mu_0 e^2 \beta_e \beta_i n_0 (u_{yi} - u_{ye})^2} \right)^{1/2},
\label{Lsymm}
\end{equation}
which is symmetric in electron and ion parameters. Using Eq. (\ref{b0luy}), expressions for $L$ using only electron or only ion parameters can be derived from Eq. (\ref{Lsymm}).

The relation of the other macroscopic parameters to the microscopic parameters are more obvious. 
Equation (\ref{b0n0}) directly relates $B_0$, the magnetic field strength in the limit $z \to \infty$, with 
$\beta_e$, $\beta_i$ and $n_0$. In the original Harris sheet case, $n_0$ is the maximum value of the $z$-dependent part of the particle density at $z=0$, and Eq. (\ref{b0n0}) simply states that the magnetic pressure for $z \to \infty$ has to be equal to the plasma pressure at $z=0$ due to force balance. As we will see later, in the force-free Harris sheet case the meaning of $n_0$ changes, but because we have effectively separated the total force-balance into two conditions for $B_x$ and $B_y$, the same condition as for the original Harris sheet still applies for the force-free Harris sheet as well.

Equation (\ref{ra}) directly shows that for the force-free Harris sheet we have $a=1/2$.

Equation (\ref{pbb}) relates the constant background pressure $P_b$ to the microscopic parameter $b$, which is representing the magnitude of the part of the DF which depends only on $H_s$. Obviously, $b$ is simply the ratio of the background pressure $P_b$ 
to the pressure $(\beta_e +\beta_i) n_0/(\beta_e \beta_i)$.
Equation (\ref{b0lux}), together with Eq. (\ref{b0luy}), allows us to relate $u_{ys}$ to $u_{xs}$ by writing
\begin{equation}
|u_{ys} |=  |u_{xs}| . \label{uxuy}
\end{equation}
%

An expression for $N(A_x, A_y)$  for the force-free Harris sheet equilibrium
which is  symmetrical in electron and ion parameters is given by
\begin{equation}
N(A_x,A_y) = n_0\left[ a \cos\left(\frac{2A_x}{A_0}\right)
+\exp\left(\frac{2A_y}{A_0}\right) + b \right]
\label{Nmicro}
\end{equation}
with
\begin{equation}
A_0 = \frac{2(\beta_e+\beta_i)}{e\beta_e\beta_i|u_{yi}-u_{ye}|},
\label{a0def}
\end{equation}
An expression for $P_{zz}$ which is symmetrical in ion and electron parameters is obtained by using (\ref{Nmicro}) in Eq. (\ref{pzzneutral}). Using the 
vector potential for the force-free Harris sheet, (\ref{ffharrisA}), we obtain for the particle density as a function of $z$, expressed using microscopic parameters,
\begin{equation}
N(z) = n_e(z) = n_i(z) = n_0 \left[ \frac{1}{2} + b\right],
\label{nofz}
\end{equation}
the pressure $P_{zz}$ is obtained by 
multiplying $N(z)$ by $(\beta_e+\beta_i)/\beta_e \beta_i$. 

The mean bulk flow velocities of each species in the $x$- and 
$y$-directions as functions of $z$, namely
\begin{eqnarray}
<v_{xs}> &=&\frac{u_{ys} \sinh(z/L)}{(\frac{1}{2}+b)\cosh^2(z/L) }, \label{vxmean} \\
<v_{ys}> &=&\frac{u_{ys}}{(\frac{1}{2}+b)\cosh^2(z/L)}, \label{vymean}
\end{eqnarray}
which gives a current density of the form
\begin{eqnarray}
j_x &=&e n_0 (u_{yi}-u_{ye}) \frac{ \sinh(z/L)}{\cosh^2(z/L) }, \label{jxmicro} \\
j_y &=&e n_0  (u_{yi}-u_{ye}) \frac{1}{\cosh^2(z/L)}, \label{jymicro}
\end{eqnarray}

The force-free parameter $\alpha(z)$ can be directly determined by using
 (\ref{Lsymm}) in (\ref{alphaffharris}) resulting in
 \begin{equation}
 \alpha(z) = \left(\frac{\mu_0e^2 \beta_e\beta_i n_0 (u_{yi}-u_{ye})^2}{2(\beta_e+\beta_i)} \right)^{1/2}
\left\{ \cosh \left[ \left(\frac{\mu_0 e^2\beta_e\beta_i n_0 (u_{yi}-u_{ye})^2}{2(\beta_e+\beta_i)} \right)^{1/2} z\right]\right\}^{-1}.
 \label{alphamicro}
 \end{equation}
 One can easily show that this is consistent with the expression for $\alpha(z)$ obtained from the current density (\ref{jxmicro}) and (\ref{jymicro}) and the magnetic field for the force-free Harris sheet (\ref{ffharrisB}), when taking into account (\ref{b0n0}).


\subsection{The number of maxima of the DF in $v_x$ and $v_y$}

\begin{figure}
\includegraphics[width=0.8\textwidth]{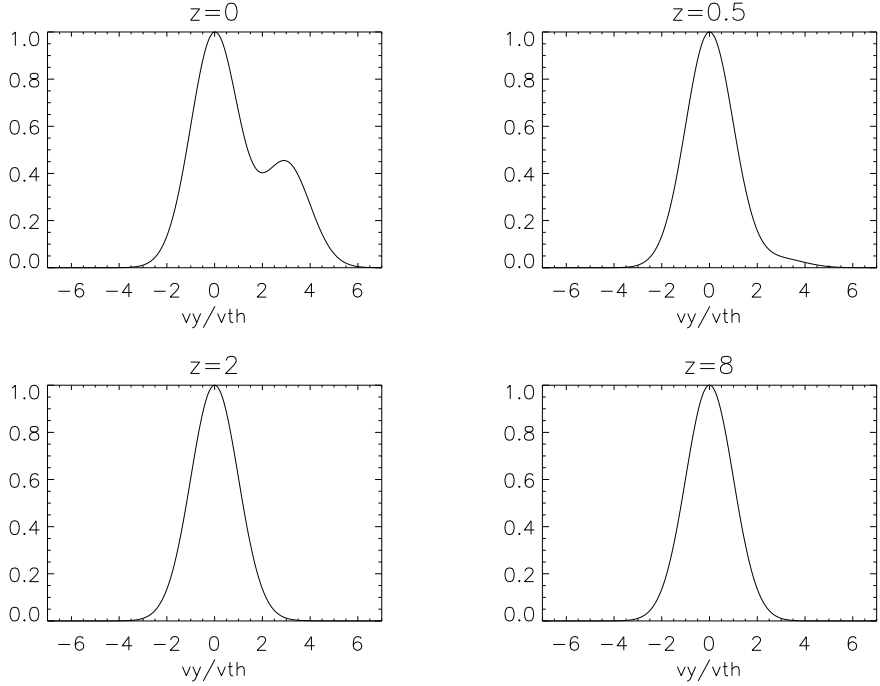}
\caption{Shape of the DF in the $v_y$-direction for various values of $z/L$ for a multiple maximum case. Here $u_{ys}= 3 v_{th,s}$, $b_s = 4.254 \cdot 10^3$ and $v_x= 0$ have been used. In the case shown here the DF has multiple maxima for small  values of $|z|$, but only a single maximum as $|z|$ increases. The value of $b_s$ has been chosen to be smaller than, but close to the critical value calculated in appendix \ref{sec:appendixa}.}
\label{fig:vydist1}
\end{figure}

\begin{figure}
\includegraphics[width=0.8\textwidth]{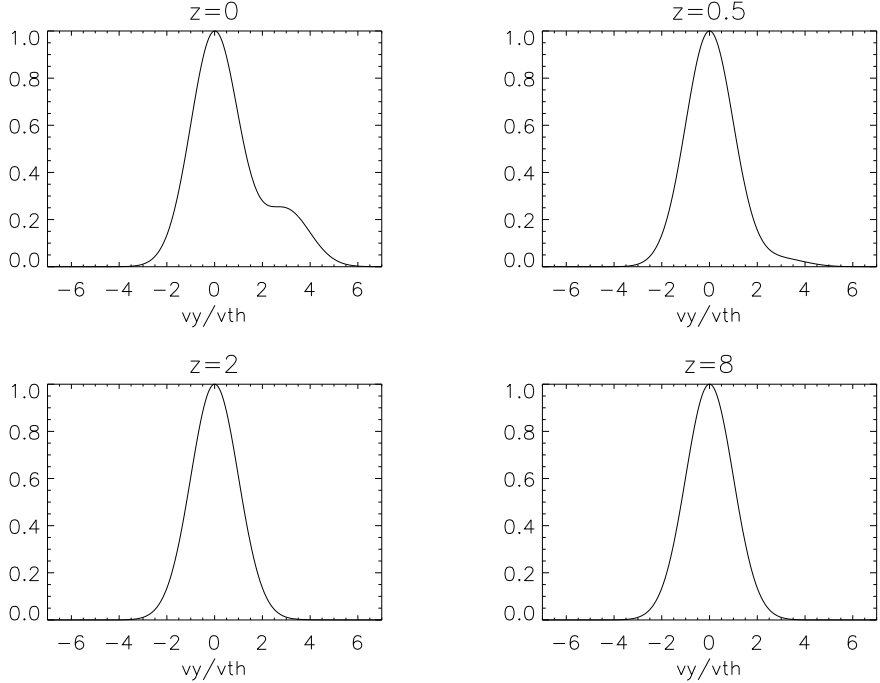}
\caption{Shape of the DF in the $v_y$-direction for various values of $z/L$ for the critical case, at which the transition between multiple maxima and a single maximum occurs. Here $u_{ys}= 3 v_{th,s}$, $b_s = 4.427 \cdot 10^3$ and $v_x= 0$ have been used. For $z=0$ the DF has a point of inflection with horizontal slope, but only one maximum. For $|z| >0$ the DF only has a single maximum. The value of $b_s$ has been chosen to be equal to the critical value calculated in appendix \ref{sec:appendixa}.}
\label{fig:vydist2}
\end{figure}

\begin{figure}
\includegraphics[width=0.8\textwidth]{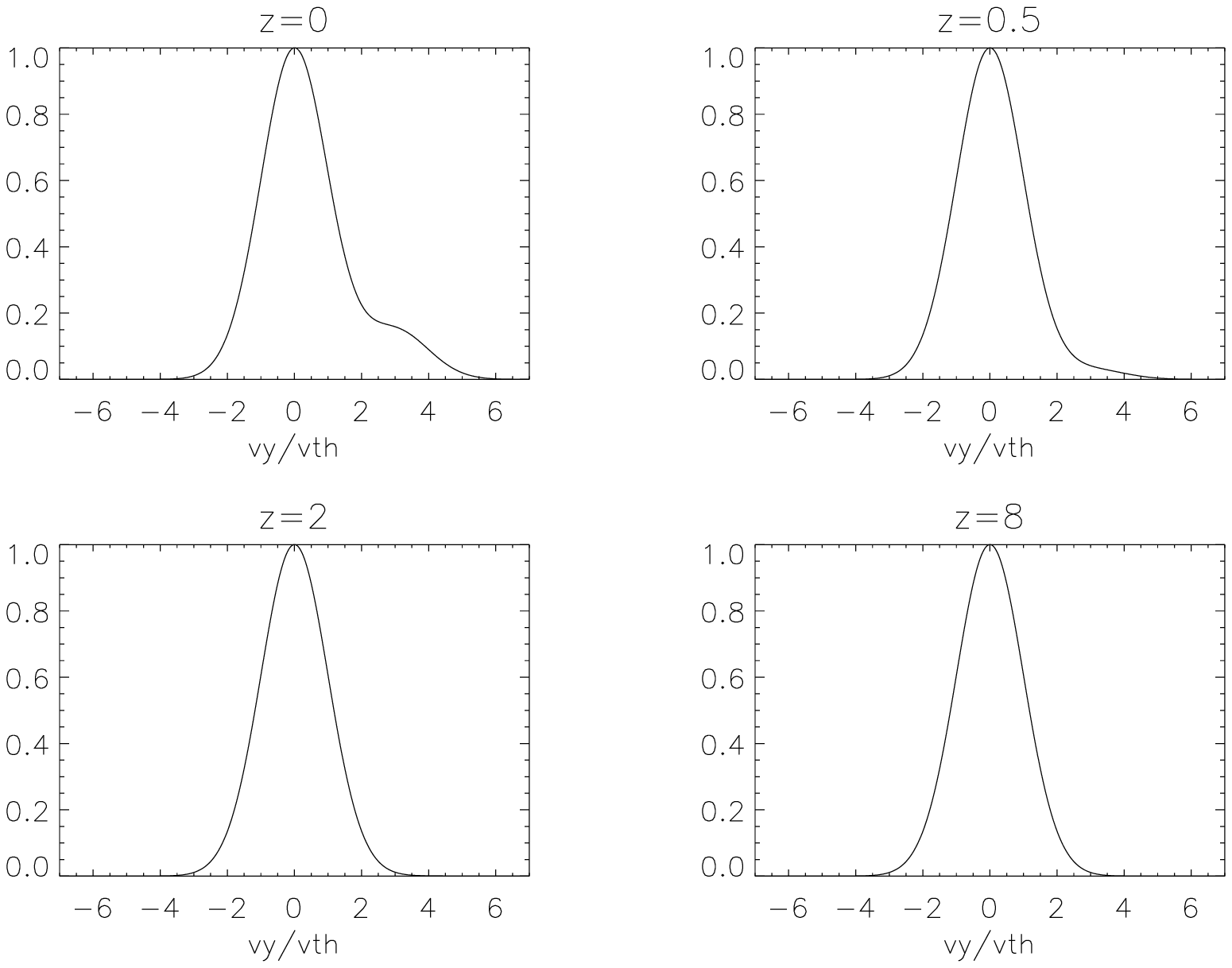}
\caption{Shape of the DF in the $v_y$-direction for various values of $z/L$ for a single maximum case. Here $u_{ys}= 3 v_{th,s}$, $b_s = 4.659 \cdot 10^3$ and $v_x= 0$ have been used. The value of $b_s$ has been chosen to be greater than the critical value calculated in appendix \ref{sec:appendixa}.}
\label{fig:vydist3}
\end{figure}

One of the interesting features of the force-free Harris sheet DF (\ref{fullDF}) is that it can have multiple maxima in both the $v_x$- and the $v_y$-directions. We shall discuss the $v_y$-direction first as it is simpler to understand. Looking at the structure of the DF in the $v_y$ direction one can immediately see that it consists of the Harris sheet DF part, which is a Maxwellian distribution function drifting with a constant velocity $u_{ys}$ in the $v_y$-direction, and a part which, if regarded purely as function of $v_y$, is Maxwellian at rest. It is intuitively clear that one should get two maxima in $v_y$ if the drift velocity $u_{ys}$ increases, because the drifting Maxwellian moves towards the tail of the Maxwellian at rest. As we show in appendix  \ref{sec:appendixa} it is relatively straightforward to work out that a necessary condition
for having more than one maximum in the $v_y$-direction is 
\begin{equation}
|u_{ys}| > 2 v_{th,s},
\label{uyconditionx}
\end{equation}
i.e. the constant drift velocity has to be larger than twice the thermal velocity. There  is, however, a second condition on the parameter $b_s$ that also needs to be satisfied for the DF to have more than one maximum in $v_y$. We derive and state the exact condition in appendix
\ref{sec:appendixa},  but its physical meaning is very easy to understand. If $b_s$ 
exceeds a certain limiting value, the part of the DF which does have vanishing average velocity in the $y$-direction dominates over the other part of the DF, so that a second maximum cannot develop even if 
(\ref{uyconditionx}) is satisfied. Usually, this condition on $b_s$ will not be very restrictive, though,  as the upper limit for $b_s$ grows exponentially with $u_{ys}^2/v_{th,s}^2$ 
(see appendix \ref{sec:appendixa}).
We show examples of DFs as functions of $v_y$ for the different cases in Figs. \ref{fig:vydist1} -
 \ref{fig:vydist3}. For these figures the values of $b_s$ have been chosen to be close to the critical value discussed in appendix \ref{sec:appendixa} for illustrative purposes. The values for $b_s$ are of the order $4.5 \cdot 10^3$ for the examples shown, which corroborates the point made above regarding the exponential growth of the limiting value.

\begin{figure}
\includegraphics[width=0.7\textwidth]{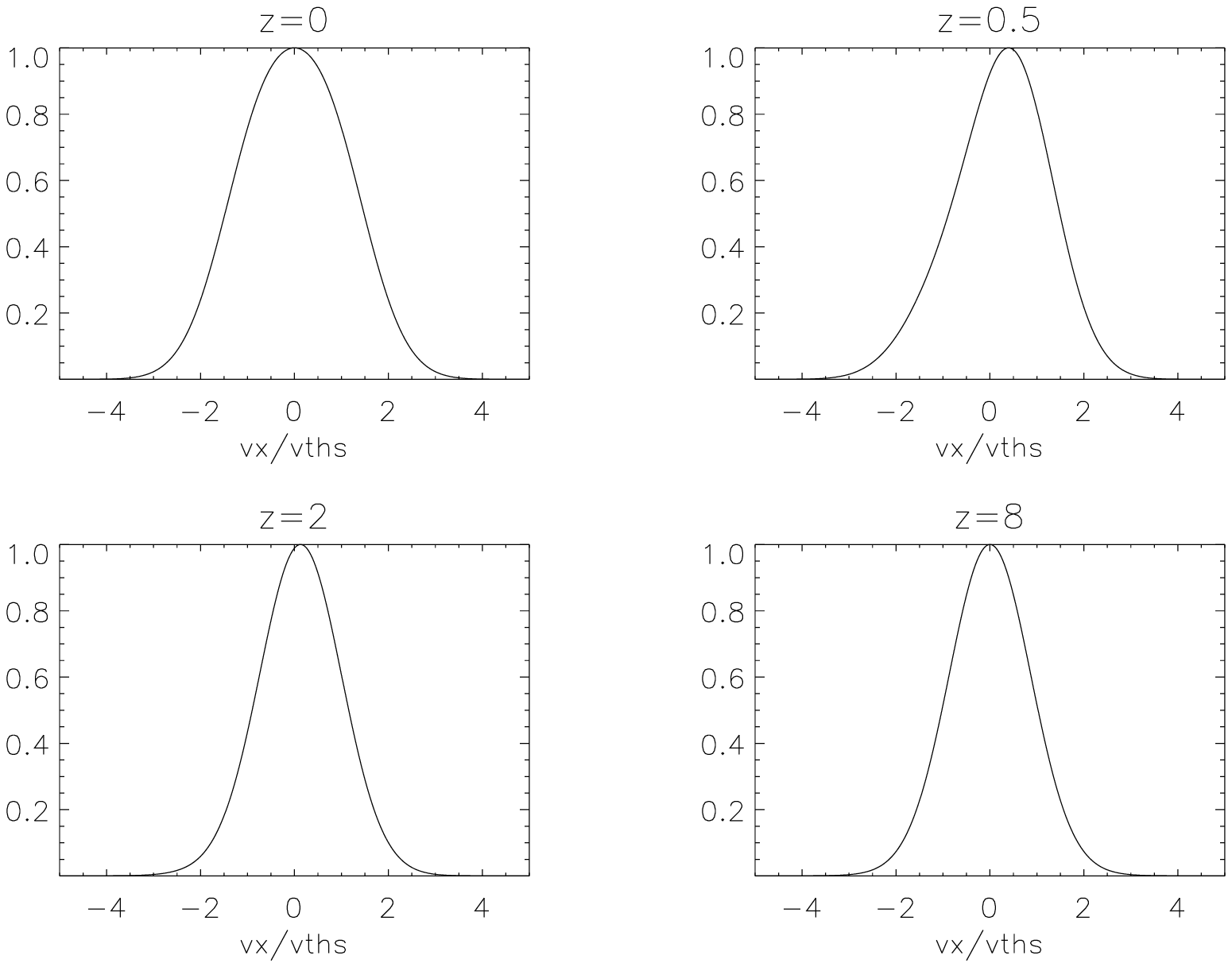}
\caption{Shape of the DF in the $v_x$-direction for various values of $z/L$ for a single maximum case. Here $u_{ys}=  v_{th,s}$, $b_s =2.85$ and $v_y= 0 $ have been used. }
\label{fig:vxdist1}
\end{figure}

\begin{figure}
\includegraphics[width=0.7\textwidth]{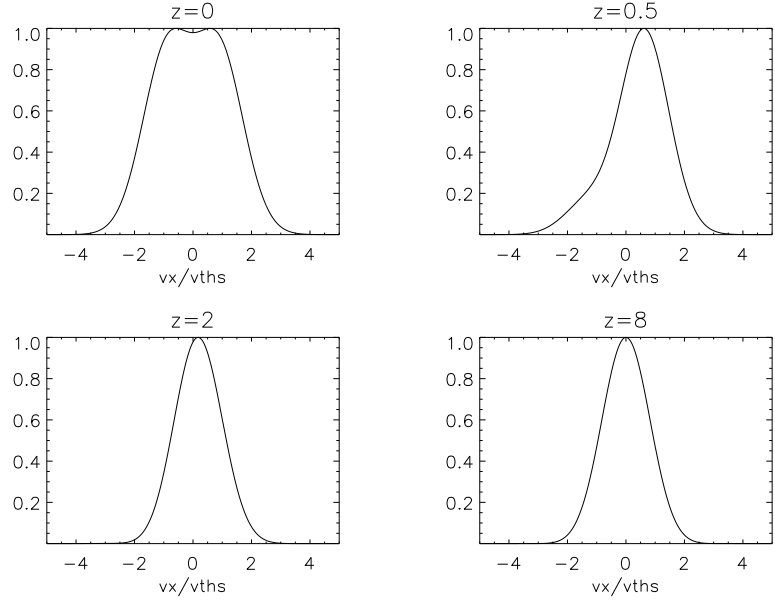}
\caption{Shape of the DF in the $v_x$-direction for various values of $z/L$ for a multiple maximum case. Here $u_{ys}=  v_{th,s}$, $b_s =1.43$ and $v_y= 0$ have been used. In the case shown here the DF has multiple maxima close to the sheet centre ($z=0$), but a single maximum for larger distances from the sheet centre.}
\label{fig:vxdist2}
\end{figure}

\begin{figure}
\includegraphics[width=0.7\textwidth]{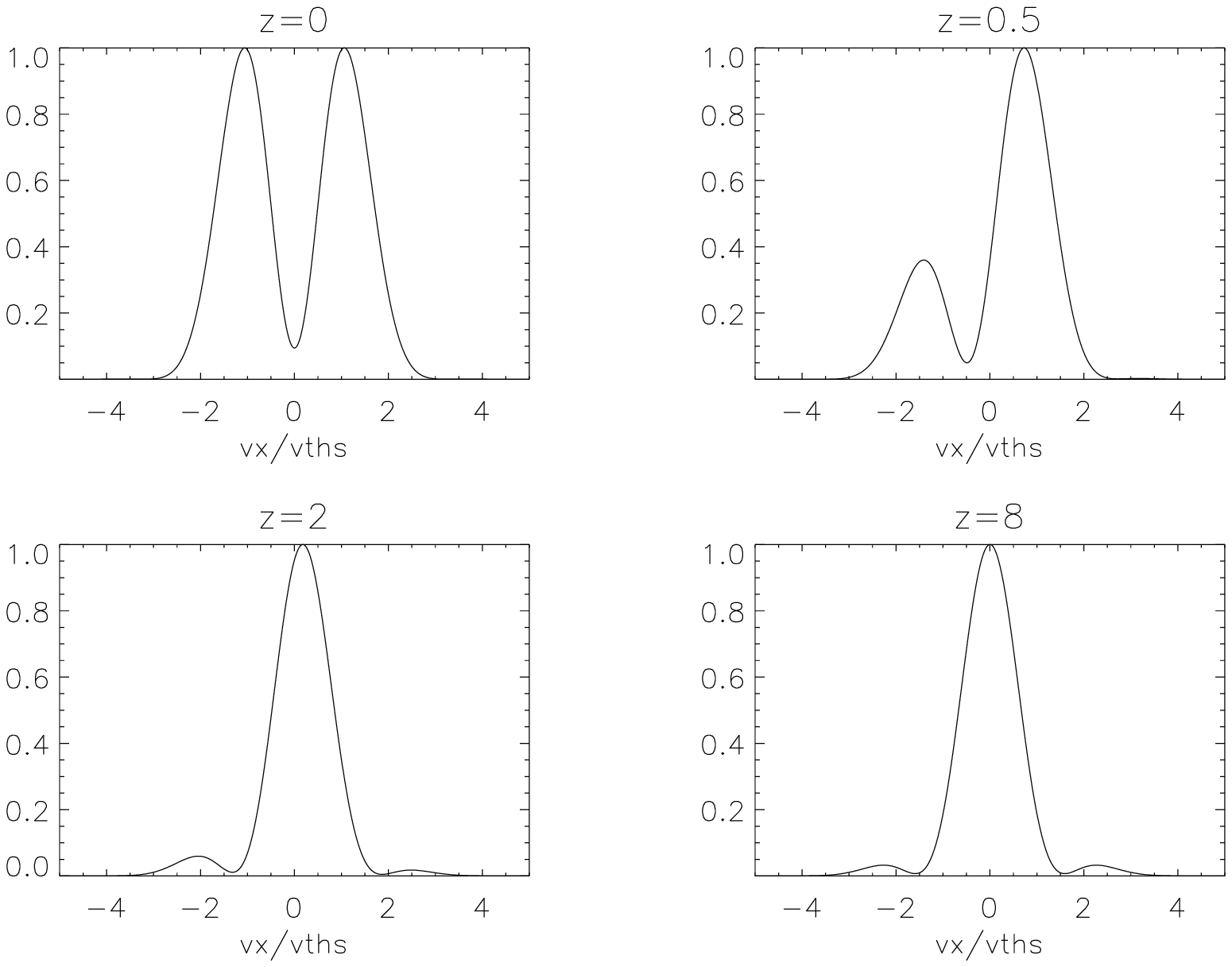}
\caption{Shape of the DF in the $v_x$-direction for various values of $z/L$ for a multiple maximum case. Here $u_{ys}= 2v_{th,s}$, $b_s = 28.66$ and $v_y=0$ have been used. In the case shown here the DF has multiple maxima for all values of $z$.}
\label{fig:vxdist3}
\end{figure}

We now turn to the dependence of the DF on $v_x$. Due to the cosine-dependence it is clear that the possibility of multiple maxima in $v_x$ exists. We discuss the details of the calculation in appendix \ref{sec:appendixb}. From the analysis in appendix  \ref{sec:appendixb} we find that the condition for having just a single maximum in $v_x$ is 
\begin{equation}
b_s > \frac{1}{2} \exp\left( \frac{u_{ys}^2}{v_{th,s}^2}\right) \left(\frac{u_{ys}^2}{v_{th,s}^2} +1\right).
\label{vxsinglecondition}
\end{equation} 
This condition on $b_s$ can be understood in the same way as the similar condition on $b_s$ derived for the $v_y$-dependence. If $b_s$ is large enough the Maxwellian background plasma it represents dominates the part of the DF with the cosine dependence and we only have a single maximum of the distribution function. If the condition (\ref{vxsinglecondition}) is not satisfied then we have multiple maxima in $v_x$, but their existence still depends on the values of $z/L$ and 
$v_y$. Obviously, for small $u_{ys}/v_{th,s}$ the limiting value on the right hand 
side of (\ref{vxsinglecondition}) is $1/2$, which is consistent with the absolute lower limit on $b_s$
mentioned before. Examples of the different cases are shown in Figs. \ref{fig:vxdist1} to \ref{fig:vxdist3}.

A slightly different perspective on the discussion above can be provided if we express the ratio 
$u_{ys}/v_{th,s}$ in terms of the current sheet thickness $L$. Using Eq. (\ref{b0luy}), we get
\begin{equation}
\frac{u_{ys}^2}{v_{th,s}^2} = 4 \frac{r_{g,s}^2}{L^2},
\end{equation}
where $r_{g,s} = m_s v_{th,s}/e B_0$ is the thermal gyroradius of species $s$. If all parameters 
except $L$ and $u_{ys}$ are fixed, it is obvious that a decrease in the current sheet thickness will eventually lead to multiple maxima in the DF, first in $v_x$ by 
violating condition (\ref{vxsinglecondition}) and then in $v_y$ as well. This may obviously have implications for possible velocity instabilities of the system, e.g. the two-stream or 
bump-on-tail instabilities, apart from macroscopic instabilities of the current sheet, e.g. the collisionless tearing mode. A detailed investigation of the stability properties of this inhomogeneous Vlasov-Maxwell equilibrium would be very interesting, but is beyond the scope of the present paper and will be left for future work.
 
\section{Summary and Conclusions}
\label{sec:conclusions}

We have given a detailed presentation of the derivation and the properties of the DF for the 
collisionless force-free Harris sheet found in Ref. \onlinecite{Harrison-2009b}. In particular, we have shown how the microscopic parameters of the DF are related to the macroscopic parameters of the magnetic field. We have also given a detailed derivation of the conditions on the parameters of the DF to ensure that it has only a single maximum in $v_x$ and in $v_y$. We have shown that as the current sheet thickness decreases the condition for multiple maxima will eventually be violated and we have suggested that this may lead to velocity space instabilities in addition to other macroscopic instabilities for thin current sheets. The stability properties of the VM equilibrium are a very interesting topic for further investigations.

\acknowledgments{
The authors acknowledge support by the UK's Science and Technology Facilities Council 
and by the European Commission through the SOLAIRE Network (MTRN-CT-2006-035484).
}

\appendix

\section{Condition for two maxima in the $v_y$-direction}

\label{sec:appendixa}

From a mathematical point of view it is easier to write the DF as a function of the momenta when carrying out
this calculation. The $p_{zs}$-dependence does not play any role in the calculation and can be integrated out. The reduced DF for $p_{xs}$ and $p_{ys}$ then reads
\begin{equation}
\bar{F}_s(\bar{z},\bar{p}_{xs},\bar{p}_{ys} )= 
\exp\left\{-\frac{1}{2 \bar{u}_{ys}^2} \left[ \left(\bar{p}_{xs} -  \bar{A}_x\right)^2 + 
                                                                       \left(\bar{p}_{ys} -  \bar{A}_y\right)^2\right] \right\}
                                                                       \left[a_s\cos(\bar{p}_{xs}) + \exp(\bar{p}_{ys}) +b_s \right],
\label{reducedDF}                                                                       
\end{equation}
where $\bar{F}_s = 2\pi (m_s v_{th,s})^2 F_s/ n_{0s}$, with $F_s = \int f_s d p_{zs}$, 
$\bar{u}_{ys}= u_{ys}/v_{th,s}$, $\bar{p}_{xs} = \beta_s u_{ys} p_{xs}$, 
$\bar{p}_{ys} = \beta_s u_{ys} p_{ys}$,  $\bar{A}_x = q_s \beta_s u_{ys} A_x = 2 A_x/(B_0 L)$ and
similarly $\bar{A}_y = q_s \beta_s u_{ys} A_y = 2 A_y/(B_0 L)$.

For an extremum of $\bar{F}_s$ in the $\bar{p}_{ys}$ direction the derivative
%
\begin{eqnarray}
\frac{\partial \bar{F}_s}{\partial \bar{p}_{ys}}  & = &
\exp\left\{-\frac{1}{2 \bar{u}_{ys}^2} \left[ \left(\bar{p}_{xs} -  \bar{A}_x\right)^2 + 
                                                                       \left(\bar{p}_{ys} -  \bar{A}_y\right)^2\right] \right\} \times
\nonumber \\
& & \mbox{\hspace{-0cm}}                                                                       
\left\{ \exp(\bar{p}_{ys}) - \frac{1}{\bar{u}_{ys}^2} \left(\bar{p}_{ys} - \bar{A}_y \right)
\left[a_s \cos(\bar{p}_{xs}) +  \exp(\bar{p}_{ys}) + b_s \right]
\right\}
\end{eqnarray}
must vanish, leading to the condition
\begin{equation}
\bar{p}_{ys} -\bar{A}_y = \frac{ \bar{u}_{ys}^2  \exp(\bar{p}_{ys})}{a_s \cos(\bar{p}_{xs}) +  \exp(\bar{p}_{ys}) + b_s }.
\label{firstderivzero}
\end{equation}
We remark that the right hand side is well-defined because $b_s > a_s \ge 0$.

The left hand side of (\ref{firstderivzero}) is a linear function of unit slope in $\bar{p}_{ys}$, which crosses the $\bar{p}_{ys}$-axis at $\bar{p}_{ys}=\bar{A}_{y}$. As $\bar{A}_{y}$ varies between $-\infty$ and 
$0$, the left hand side intercepts the $\bar{p}_{ys}$-axis for negative values of  $\bar{p}_{ys}$. The right hand side of  (\ref{firstderivzero})  can be rewritten as
\begin{equation}
R(\bar{p}_{ys}) =\frac{A}{ 1+B\exp(-\bar{p}_{ys} )} ,
\label{RHSysimple}
\end{equation}
where $A=\bar{u}_{ys}^2 >0$ and $B= a_s \cos(\bar{p}_{xs}) +b_s >0$. The function 
(\ref{RHSysimple}) is positive, increases monotonically and is bounded between $0$ and $A$. Therefore, a necessary condition for multiple maxima of the DF in  $v_y$ (or $p_{ys}$) is that the maximum slope of $R(\bar{p}_{ys})$ must be larger than $1$. Otherwise the (\ref{firstderivzero}) can
only have a single solution, implying a single maximum for the distribution function. It is straightforward to show that $R(\bar{p}_{ys})$ has its maximum slope $A/4$ at 
$\bar{p}_{ys} = \ln B$. So the necessary condition for multiple maxima is $A/4 >1$ which translates
into
\begin{equation}
|u_{ys}| > 2 v_{th,s}.
\label{vy_nec_cond}
\end{equation}

\begin{figure}
\includegraphics[width=0.7\textwidth]{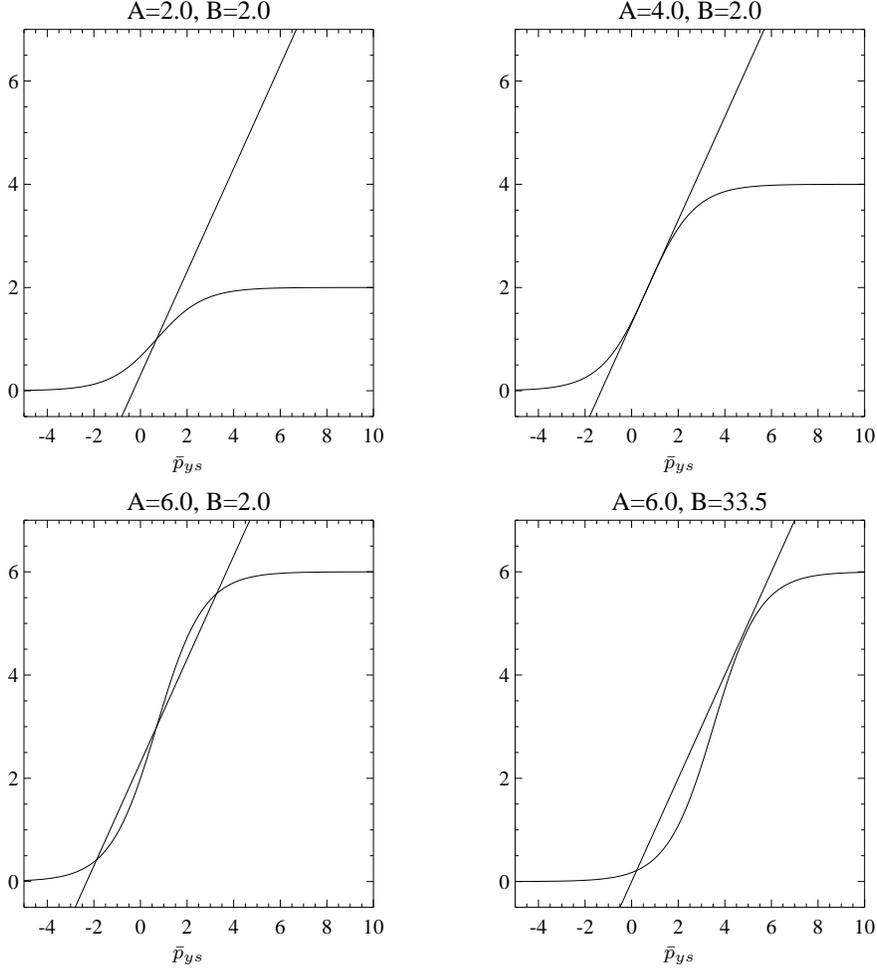}
 \caption{Upper left panel: A case for which $R(\bar{p}_{ys})$ has a maximum slope smaller than one ($A=2.0$, $B=2.0$). Upper right panel: A case for which $R(\bar{p}_{ys})$ has a maximum slope equal to one ($A=4.0$, $B=2.0$). Lower left panel: 
 A case for which $R(\bar{p}_{ys})$ has a maximum slope larger than one ($A=6.0$, $B=2.0$).
 A case for which $R(\bar{p}_{ys})$ has a maximum slope greater than one, but
 for which $B$ is larger than $B_l$ ($A=6.0$, $B=33.5$, $B_l=30.42$). The straight line shown passes through the point of maximum slope in all plots apart from the lower right panel. In the lower right panel the straight line passing through the origin is shown.}
\label{fig:ycondition}
\end{figure}

However, (\ref{vy_nec_cond}) is not sufficient, because even if it is satisfied, it is still possible that
$\bar{p}_{ys} -\bar{A}_y$ intersects with $R(\bar{p}_{ys})$ only once, namely if the value of $B$ is large enough. As discussed above the left hand side of (\ref{firstderivzero}) can only cross the
$\bar{p}_{ys}$-axis for $\bar{p}_{ys} \le 0$, depending on the value of 
$\bar{A}_y$ (and thus $z/L$). Since the $\ln B$ is positive it can happen that $R(\bar{p}_{ys})$ takes on its maximum slope too far to the right for more than one intersection between the two functions to happen.The transition between three intersections to one intersection happens 
at the value of $B$ for which the straight line of slope one through the origin just touches the graph of $R(\bar{p}_{ys})$ at the point where it also has unit slope (see Fig. \ref{fig:ycondition}). 
One can easily calculate the value of $\bar{p}_{ys}$ for which the function $R(\bar{p}_{ys})$ has unit slope as
\begin{equation}
\bar{p}_{ys,u} = \ln(2B) - \ln(A-2-\sqrt{A(A-4)}).
\end{equation}
Two remarks are to be made here:
\begin{itemize}
\item
$\bar{p}_{ys,u}$ only has a real value if $A>4$, which is consistent with the condition  found before for $R(\bar{p}_{ys})$ to have slope greater than unity anywhere;

\item For $A>4$, the function $R(\bar{p}_{ys})$ has unit slope at two values of $\bar{p}_{ys}$, of which one has to choose the larger one (see Fig. \ref{fig:ycondition}), as we have done above. 

\end{itemize}

The limiting value for $B$ can now derived from 
\begin{equation}
\bar{p}_{ys,u} = R(\bar{p}_{ys,u})
\end{equation}
leading to
\begin{equation}
B_l = \frac{1}{2} [A-2 -\sqrt{A(A-4)}]\exp\left(\frac{2A}{A-\sqrt{A(A-4)}} \right),
\label{defBl}
\end{equation}
so the sought for condition is
\begin{equation}
B < B_l.
\end{equation}
Since $B$ still depends upon $\bar{p}_{xs}$ we have to replace it by the minimum value it can take on as function of $\bar{p}_{xs}$ to get a condition which is independent of $\bar{p}_{xs}$.

In summary, the DF has more than one maximum in $p_{ys}$ (and thus in $v_y$) if the following conditions are both satisfied
\begin{eqnarray}
|u_{ys}| & > & 2 v_{th,s} , \label{vxcond1}\\
 b_s  &< &\frac{1}{2v_{th,s}} (u_{ys}^2 - 2 v_{th,s}^2-|u_{ys}| \sqrt{u_{ys}^2 -4v_{th,s}^2}) 
 \exp\left( \frac{2 u_{ys}^2}{u_{ys}^2 - |u_{ys}| \sqrt{u_{ys}^2 -4v_{th,s}^2}} \right) \nonumber\\
 & & \mbox{\hspace{1cm}} +
 \frac{1}{2}\exp\left(\frac{u_{ys}^2}{v_{th,s}^2} \right),
 \label{vxcond2}
\end{eqnarray}
where we have made use of (\ref{defa}) to replace $a_s$.

\section{Condition for multiple maxima in the $v_x$-direction}

\label{sec:appendixb}

\begin{figure}
\includegraphics[width=0.9\textwidth]{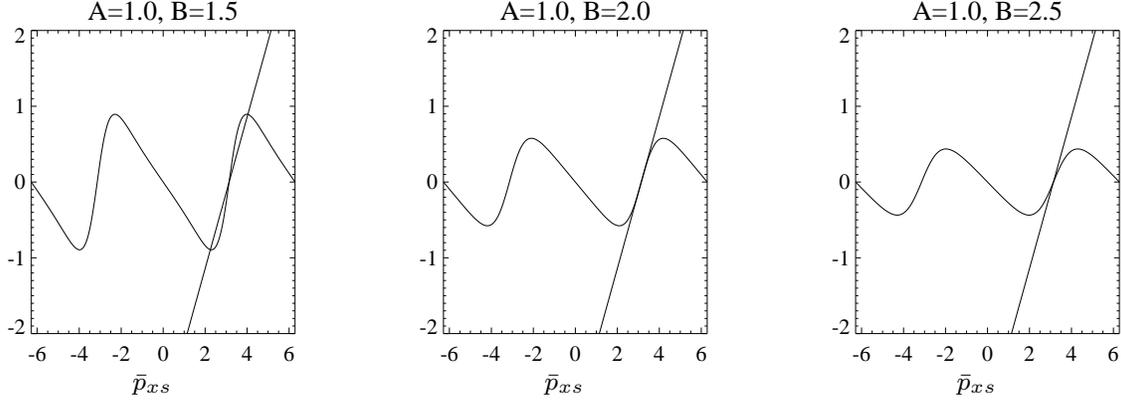}
 \caption{Left panel: A case in which $R(\bar{p}_{xs})$ has a maximum slope greater than unity ($A=1.0$, $B=1.5$); Middle panel: The limiting case with maximum slope equal unity ($A=1.0$, $B=2.0$); Right panel: A case in which $R(\bar{p}_{xs})$ has a maximum slope less than unity ($A=1.0$, $B=2.5$). For these plots the straight line of unit slope has been chosen to cross the $\bar{p}_{xs}$-axis at $\bar{p}_{xs}=\pi$.}
\label{fig:xcondition}
\end{figure}

The analysis here is very similar to that in appendix \ref{sec:appendixa}. Again we use the reduced DF (\ref{reducedDF}) expressed as a function of the canonical momenta $\bar{p}_{xs}$ and
$\bar{p}_{ys}$. Taking the derivative of $\bar{F}_s$ with respect to $\bar{p}_{xs}$ gives
\begin{eqnarray}
\frac{\partial \bar{F}_s}{\partial \bar{p}_{xs}}  & = &
-\exp\left\{-\frac{1}{2 \bar{u}_{ys}^2} \left[ \left(\bar{p}_{xs} -  \bar{A}_x\right)^2 + 
                                                                       \left(\bar{p}_{ys} -  \bar{A}_y\right)^2\right] \right\} \times
\nonumber \\
& & \mbox{\hspace{-0cm}}                                                                       
\left\{ a_s\sin(\bar{p}_{xs}) + \frac{1}{\bar{u}_{ys}^2} \left(\bar{p}_{xs} - \bar{A}_x \right)
\left[a_s \cos(\bar{p}_{xs}) +  \exp(\bar{p}_{ys}) + b_s \right]
\right\}.
\end{eqnarray}
Setting this to zero gives the equation
\begin{equation}
\bar{p}_{xs} - \bar{A}_x = - \frac{ \bar{u}_{ys}^2 a_s \sin(\bar{p}_{xs})}{
a_s \cos(\bar{p}_{xs}) +  \exp(\bar{p}_{ys}) + b_s},
\end{equation}
or, in an abbreviated form
\begin{equation}
\bar{p}_{xs} - \bar{A}_x = R(\bar{p}_{xs}),
\label{firstderivxzero}
\end{equation}
with
\begin{equation}
R(\bar{p}_{xs}) = - \frac{C \sin(\bar{p}_{xs})}{\cos(\bar{p}_{xs}) + D}
\label{RHSxsimple}
\end{equation}
where $C= \bar{u}_{ys}^2>0$ and $D= (b_s +  \exp(\bar{p}_{ys}))/a_s>1$, because $b_s > a_s$.
The function (\ref{RHSxsimple}) is a bounded periodic function of $\bar{p}_{xs}$. 
Furthermore, $\bar{A}_x = 4 \arctan(\exp(z/L))$ varies between $0$ and $2\pi$, so the left hand side of (\ref{firstderivxzero}) can only cross the $\bar{p}_{xs}$-axis between $0$ and $2\pi$. The slope of $R(\bar{p}_{xs})$ is given by
\begin{equation}
\frac{\partial R}{\partial \bar{p}_{xs}} = -C\frac{D \cos(\bar{p}_{xs}) +1}{( \cos(\bar{p}_{xs})+D)^2 },
\label{slopeRx}
\end{equation}
which shows that $R(\bar{p}_{xs})$ has a positive slope for $\cos(\bar{p}_{xs}) < -1/D$, which is always satisfied for some $\bar{p}_{xs}$ in the interval $0\le \bar{p}_{xs} \le 2\pi$. Therefore, a necessary and sufficient condition for
multiple maxima of the DF in $v_x$ is that $R(\bar{p}_{xs})$ has a maximum slope which is larger than unity. Examples for the different cases are shown in Fig. \ref{fig:xcondition}.

Taking the derivative of (\ref{slopeRx}) we get
\begin{equation}
\frac{\partial^2 R}{\partial \bar{p}_{xs}^2} =C \sin(\bar{p}_{xs})
\frac{D^2-2 -D\cos(\bar{p}_{xs}) }{( \cos(\bar{p}_{xs})+D)^3}.
\end{equation}
A brief calculation shows that $R(\bar{p}_{xs})$ has positive slope only for $\bar{p}_{xs}=n \pi$ with 
$n$ an odd integer. The maximum value of the slope is given by $C/(D-1)$, which leads to the condition
\begin{equation}
C < D-1
\end{equation}
for the DF to have only one maximum. The lowest value $D$ can 
take (as a function of $\bar{p}_{ys}$) is $D= b_s/a_s$ so that we finally arrive at the condition
\begin{equation}
b_s > \frac{1}{2}\exp\left(\frac{u_{ys}^2}{v_{th,s}^2} \right)\left( \frac{u_{ys}^2}{v_{th,s}^2} +1 \right),
\end{equation}
for the DF to have only one maximum in $v_x$, where we have used (\ref{defa}) and (\ref{ra}) to replace $a_s$.

 \bibliographystyle{apsrev}


\end{document}